\documentclass[12pt,journal,onecolumn,draftclsnofoot]{IEEEtran}
\usepackage{amsmath}
\usepackage{amssymb}
\usepackage{graphicx}
\usepackage{color}
\usepackage{lineno}
\usepackage{cite}
\usepackage{lettrine}
\renewcommand\IEEEkeywordsname{Index Terms}

\IEEEoverridecommandlockouts

\begin{document}

\title{\Huge Adaptive Multiuser MCDM for Underwater Acoustic Communications} \vspace{-8pt}
\author{Song-Wen Huang and Dimitris A. Pados
\vspace{2pt}
\thanks{Song-Wen Huang is with Autonomygo, Inc., Santa Clara, CA 95050, USA. Dimitris A. Pados is with the I-SENSE, Department of Computer and Electrical Engineering and Computer Science, Florida Atlantic University, Boca Raton, FL 33431. (e-mail: songwenh@buffalo.edu; dpados@fau.edu)}
\vspace{-25pt}}

\maketitle

\begin{abstract}
Chirp signals have been exploited extensively in radar and underwater acoustic (UW-A) communication systems for their robustness to multipath and superior correlation properties. We present a multiuser multicarrier chirp-division multiplexing (MU-MCDM) system which orthogonal chirp spread spectrum are utilized in frequency subcarriers for multiuser applications. Furthermore, orthogonal chirp transform (OCT) and discrete orthogonal chirp transform (DOCT) are developed for continuous and discrete implementations of MCDM systems. In addition, we design a maximum likelihood (ML) receiver capable of synchronization, channel estimation, and symbol detection. MU-MCDM systems consider factors, including preamble length, number of subcarriers, portion of pilots, guard period, and multiuser access scenarios to adapt time-variant UW-A channels. The bit-error-rate (BER) performance and adaptive transmission are evaluated in simulation studies. Therefore, we have demonstrated MCDM modulation systems in simulations that they improve performance of UW-A communications adaptively and support multiple users. Moreover, MCDM architectures can be applied in higher order modulations for providing higher transmission rates in UW-A channels.
\vspace{5pt}
\end{abstract}




\section{Introduction}
\label{S1}

Underwater acoustic (UW-A) communication has been developed for various applications, including offshore remote control, undersea pollution control, environmental monitoring, underwater communication and sensor networks,  surveillance and defense \cite{melodia13,demirors15,akyildiz04,stojanovic96}. 

Currently, there are still many challenges in UW-A channels, such as severe attenuation loss, multipath propagation delay, ambient noise, Doppler effect, and frequency selective fading that restrict our technology from advancing significantly \cite{melodia13}.

Chirp signals have been utilized widely in radar and sonar applications \cite{fitzgerald74,lurton02} primarily due to their resilience to multipath propagation delays and Doppler spread as well as superior correlation characteristics. Applications in UW-A communications include linear chirp signals for navigation and underwater probing \cite{austin94,chiu15}, packet synchronization \cite{sharif00,kilfoyle00}, channel estimation \cite{wu12}, reliable UW-A communications \cite{he09}, Doppler estimation and compensation \cite{lurton02,sharif00}, and robust feedback communications \cite{demirors14}.

However, linear chirps are not orthogonal rigorously \cite{springer00}. To tackle this problem, some researchers adopt various categories of orthogonal chirp transforms, such as fractional Fourier transform \cite{en14,solyman12}, and Fresnel transform \cite{ouyang16} for supporting multicarrier communications and high data rates. However, the disadvantages of these chirp transforms are high weighted computation and complex implementations. Hence, there is of interests to design orthogonal chirp waveforms and low complexity receivers in UW-A environments \cite{huang17}.

Propelled by the need for attaining higher data rates and multiple users, chirp signals have been utilized in multicarrier communications. Some researchers applied chirps in OFDM for enhancing Doppler spread estimation, and delay tolerance \cite{dida16,barbarossa01}. Some focus on waveform designs for multiple transmitters in radar systems \cite{kim13}, orthogonal chirp signals in multiple input multiple output (MIMO) systems \cite{wang15}, and CSS for enhancing orthogonality of transmission of different users \cite{yang11,cheng15}. Several works are built on different categories of chirp bases in multicarrier systems, e.g., fractional Fourier transform (FrFT), fractional cosine transform, and Fresnel chirp transform \cite{en04,solyman12,attar17,ouyang16}. On the other hand, chirps can be exploited in multiuser communications for multiple access \cite{shen06_div}. Multicode scheme can enable simultaneous transmission of multiple chirp waveforms \cite{liu06}. However, these chirp transforms are of high computational overhead and complex implementations. Hence, developing an efficient and low complexity multicarrier chirp system is still of interests and researches in RF wireless communications.

In this paper, we consider a single-input single-output (SISO) UW-A communications that information symbols are carried over orthogonal chirp frequency subcarriers in adaptive multiuser multicarrier chirp-division multiplexing (MU-MCDM) systems. The orthogonality of chirp waveforms are analyzed by their cross-correlation coefficients. Both continuous orthogonal chirp transform (OCT) and discrete orthogonal chirp transform (DOCT) are introduced for time-domain or digital implementations. The system is adaptive in preamble duration, number of subcarriers, portion of pilots, guard interval, and multiuser scenario to optimize performance of transmission in variant underwater environments. Moreover, we develop a low complexity coherent receiver of channel estimation and maximum-likelihood (ML) symbol detection. On the other hand, we discuss the relationships of OCT and DOCT and their various applications. In addition, MCDM system is compatible with prevailing OFDM architectures. Then, simulation studies are demonstrated over UW-A multipath channels for evaluating adaptive MU-MCDM systems.

\vspace{2pt}
The rest of the paper is organized as follows. Section \ref{S2} addresses OCT representations. System Model is described in Section. \ref{S3}. Receiver design of the ML coherent receiver is introduced in Section \ref{S4}. Discussion of proposed approaches are illustrated in Section \ref{S5}. Simulation results are demonstrated in Section \ref{S6}. Finally, several concluding remarks are drawn in Section \ref{S7}.

\section{Orthogonal Chirp Waveform Design}
\label{S2}
\subsection{Chirp Waveform Design}
A linear chirp signal is considered, where waveform's frequency change linearly over time. The $m$-th chirp waveform can be represented as
\begin{align}
\psi_m(t) =  \sqrt{\frac{1}{T}} e^{j(2\pi m\Delta ft+\pi \mu t^2)},~0\leq t\leq T
\label{psi}
\end{align}
where $T$ is the symbol duration, $\Delta f$ the frequency spacing between chirps, $\mu\overset{\Delta}{=}\frac{B_c}{T}$ the chirp rate, $B_c$ the bandwidth of a chirp signal, and the waveform's energy is normalized to 1. $m\Delta f$ is the initial frequency of the $m$-th chirp waveform. For the chirp rate, $\mu >$ 0 is a up-chirp signal (frequency growing with time), while $\mu <$ 0 is a down-chirp signal (frequency reducing with time). 

Proposed chirp signals are complex, comprising both in-phase and quadrature-phase parts, so as we process channel estimation, channel coefficients should be treated as complex representations. Moreover, additional bandwidth assigned to chirp signals can support a degree of freedom for resilience of multipath effect and channel fading. Then, we examine the orthogonality of chirp waveforms in derivations.

\subsection{Orthogonality of Chirp Waveforms}
Characteristics of orthogonality of chirp waveforms are crucial in our waveform designs, so we analyze the orthogonality of chirp waveforms with cross-correlation coefficients. Then, the cross-correlation coefficient of the $k$-th and the $l$-th chirp waveforms can be derived as
\begin{align}
\begin{aligned}
\rho_{kl}  &= \int_0^{T} \psi_{k}(t) \psi_{l}^*(t)dt\\
      &= \frac{1}{T} \int_0^{T} e^{j2\pi (k-l)\Delta ft}dt\\
      &= \delta_{kl} = \left\{\begin{tabular}{cc}
      0, &\it{k $\neq$ l}\\
      1, &\it{k = l}\\
      \end{tabular}\right.
\end{aligned}
\label{cross_corre}
\end{align}
where $\Delta f \overset{\Delta}{=} \frac{1}{T}$ for orthogonal ensurance.

In (\ref{cross_corre}), cross-correlation coefficients for different chirp signals are all zeros, in other words, these chirp waveforms are orthogonal to each other. Hence, we have derived cross-correlation coefficients explicitly in closed-form expressions and proposed chirp waveform design is orthogonal.

\subsection{Orthogonal Chirp Transform}
An orthogonal chirp transform (OCT) is designed on the basis of orthogonal chirp waveforms described as
\begin{align}
\label{OCT}
X(f) =  \sqrt{\frac{1}{T}} \int_{-\infty}^{\infty} x(t)e^{-j(2\pi ft+\pi \mu t^2)}dt,
\end{align}
where $x(t)$ is a continuous-time function and $X(f)$ is its OCT representation in frequency domain. Consequently, OCT can transform a time function into its frequency form.

On the contrary, the inverse orthogonal chirp transform (IOCT) can be written as
\begin{align}
\label{IOCT}
x(t) =  \sqrt{\frac{1}{T}} \int_{-\infty}^{\infty} X(f)e^{j(2\pi ft+\pi \mu t^2)}df.
\end{align}

IOCT can transform a function in frequency into its time-domain representation. Therefore, OCT and IOCT can transform a time-domain form and its frequency representation correspondingly. OCT and IOCT can let our implementations and representations in time domain much more convenient. However, there is still of need for digital or discrete applications. Hence, we should develop their correspondent discrete transforms.

\subsection{Discrete Orthogonal Chirp Transform}
In current implementations, digital forms are more preferable, since it's impractical to store and to process in continuous-time manner, so we develop discrete forms of digital chirp transforms. Therefore, discrete orthogonal chirp transform (DOCT) can be expressed as 
\begin{align}
\label{DOCT}
X[k] = \sqrt{\frac{1}{N}} \sum_{n=0}^{N-1} x[n]e^{-j(2\pi kn+\pi \mu n^2)},
\end{align}
where $N$ is the number of OCT points, and $x[n]$ is the $n$-th sample in time. Hence, DOCT can transform digital time signals into frequency forms. 

Conversely, inverse discrete orthogonal chirp transform (IDOCT) can be described as
\begin{align}
\label{IDOCT}
x[n] = x(t)|_{t = n T_s}=  \sqrt{\frac{1}{N}} \sum_{k=0}^{N-1} X[k]e^{j(2\pi kn+\pi \mu n^2)},
\end{align}
where $T_s$ is the sampling period. Therefore, DOCT and IDOCT can transform digital signals between time and frequency domains, whereas these two representations compose a DOCT pair. 

In practical implementations or signal processing, discrete representations are more demanded, so that's the motivation for our designs for DOCT, which is effective for discrete signal processing in communication systems.

\section{System Model}
\label{S3}

\subsection{MCDM System}
Orthogonal chirp waveforms are utilized in frequency subcarriers in MCDM modulation systems. Hence, the system is benefited by the merits of chirp signals, such as robust to multipath delays and Doppler effect. In addition, it is implemented in multicarrier and multiuser architectures, so MCDM systems can support higher data rates and simultaneous multiuser access. On the other hand, they can be implemented in OCT and DOCT practices.

As a result, passband signals in MU-MCDM systems can be described as
\begin{align}
\label{s_pass}
x(t) = \sqrt{E}\sum_{k=0}^{K-1} s[k]\psi_k(t) e^{j2\pi f_c t},~0\leq t\leq T
\end{align}
where $E$ is the signal energy, $s[k]$ the $k$-th modulated symbol, $K$ the total number of frequency subcarriers, $\Delta f$ the frequency spacing between subcarriers, $\mu$ the chirp rate, $f_c$ the carrier frequency, and $T$ is the symbol period. $s[k]$ can be a symbol of any modulations, e.g. BPSK, QPSK, and 32-QAM, and its symbol energy is normalized to 1. Hence, signal energy is merely dependent on parameter $E$.

\begin{figure}
 \centering
 \includegraphics[width=0.8\textwidth]{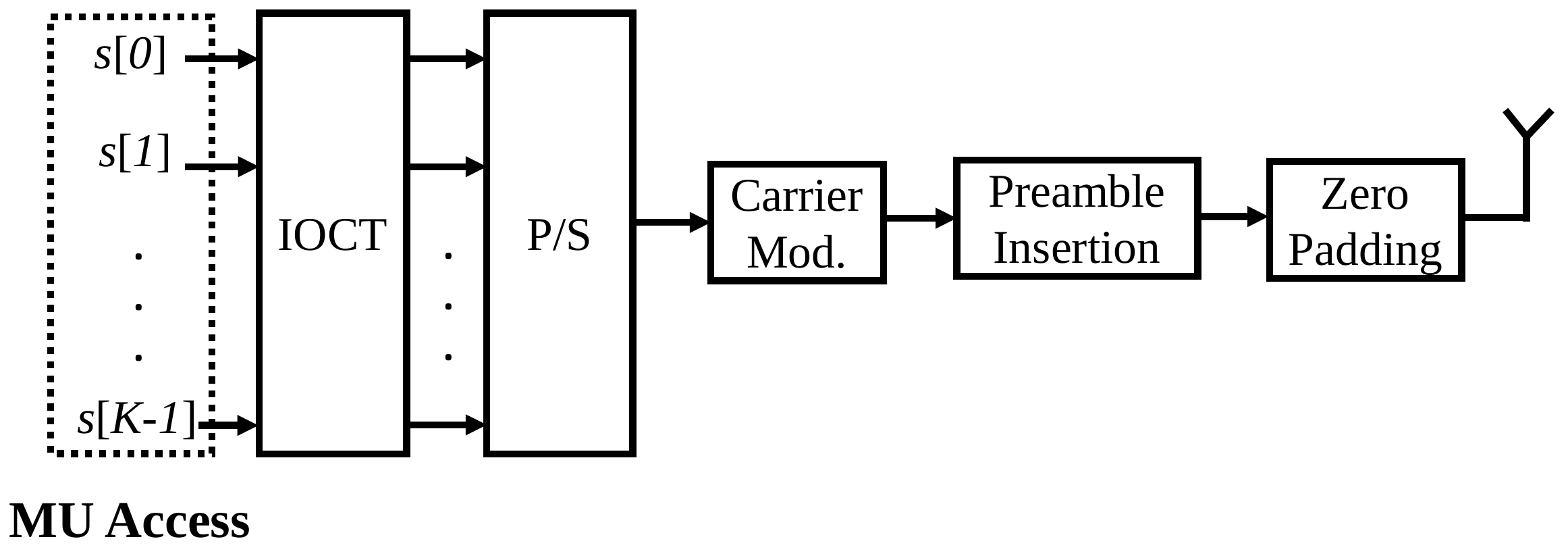} 
  \caption{MU-MCDM transmitter system model.}
  \label{Sys_Tx}
\end{figure}

Signal representations in (\ref{s_pass}) can be written as the summation of output of IOCT and carrier frequency modulation as in Fig. \ref{Sys_Tx}. The transmitted symbol $s[k]$ is modulated in frequency domain, and after exploiting IOCT, the signals are continuous-time signals, which are prepared to be transmitted in UW-A channels.

Underwater channels are modeled as multipath, independent, and time-variant represented as
\begin{align}
h(t) \overset{\triangle}{=} \sum_{m=1}^{M} h_m(t) \delta(t - \tau_m(t)),
\label{eq:3}
\end{align}
where $M$ is the total number of solvable paths, $h_m(t)$ the $m$-th path's amplitude, and $\tau_m(t)$ is the $m$-th path's delay. 

Each path's amplitude and delay are assumed to be time-invariant in a MCDM symbol period. Hence, received signals after carrier demodulation can be described as
\begin{align}
r(t) = \sqrt{E} \sum_{k=0}^{K-1}\sum_{m=1}^{M} {\tilde h}_m s[k]\psi_k(t- \tau_m) + n(t),
\label{eq:4}
\end{align}
where ${\tilde h}_m \overset{\Delta}{=} h_me^{-j2\pi f_c \tau_m} \in \mathbb{C}$ is the $m$-th path's effective channel coefficient, $\tau_m(t)$ the $m$-th path's delay, and $n(t)$ is ambient noise.

\begin{figure}
 \centering
 \includegraphics[width=0.8\textwidth]{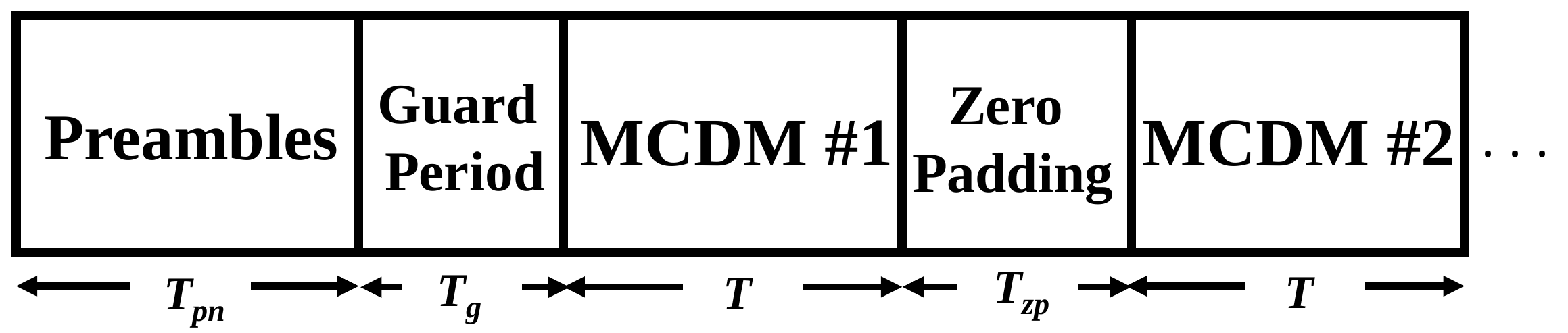} 
  \caption{MU-MCDM transmitted packet structure.}
  \label{packet}
\end{figure}

The transmitted packet structure of a MU-MCDM system is shown in Fig \ref{packet}. There is a preamble block comprising pseudo random antipodal bits $\in \{\pm 1\}^{N_{pn}}$ in the beginning of the transmitted structure for packet synchronization. A MCDM symbol contains multiuser chirp signals in a time duration $T$, and a guard period $T_g$ is inserted between preambles and a MCDM symbol. In addition, zeros are padded after a MCDM symbol with a duration $T_{zp}$, so we consider a ZP-MCDM architecture. The ZP procedure can avoid multipath delays and inter symbol interference (ISI) in UW-A channels.

Our problem objective is to design a coherent receiver with procedures, including synchronization, channel estimation, and symbol detection. Packet synchronization is proceeded by known preambles. Channel state information (CSI) is estimated by pilot symbols among frequency subcarriers. The symbol detection is conducted by estimated channel coefficients and ML symbol detections.

\subsection{Multiuser Access}
Multiuser (MU) access is the practices that enabling multiple users' simultaneous transmission in communication systems. We implement MU access by subcarrier allocation in frequency domain, so multiple users are assigned for different subcarriers and they can transmit their own information symbols in MCDM systems. There are various kinds of MU access scenarios and we focus on block-type and comb-type subcarrier allocation for multiple users.

\subsubsection{Block-type MU Access}
\begin{figure}
 \centering
 \includegraphics[width=0.8\textwidth]{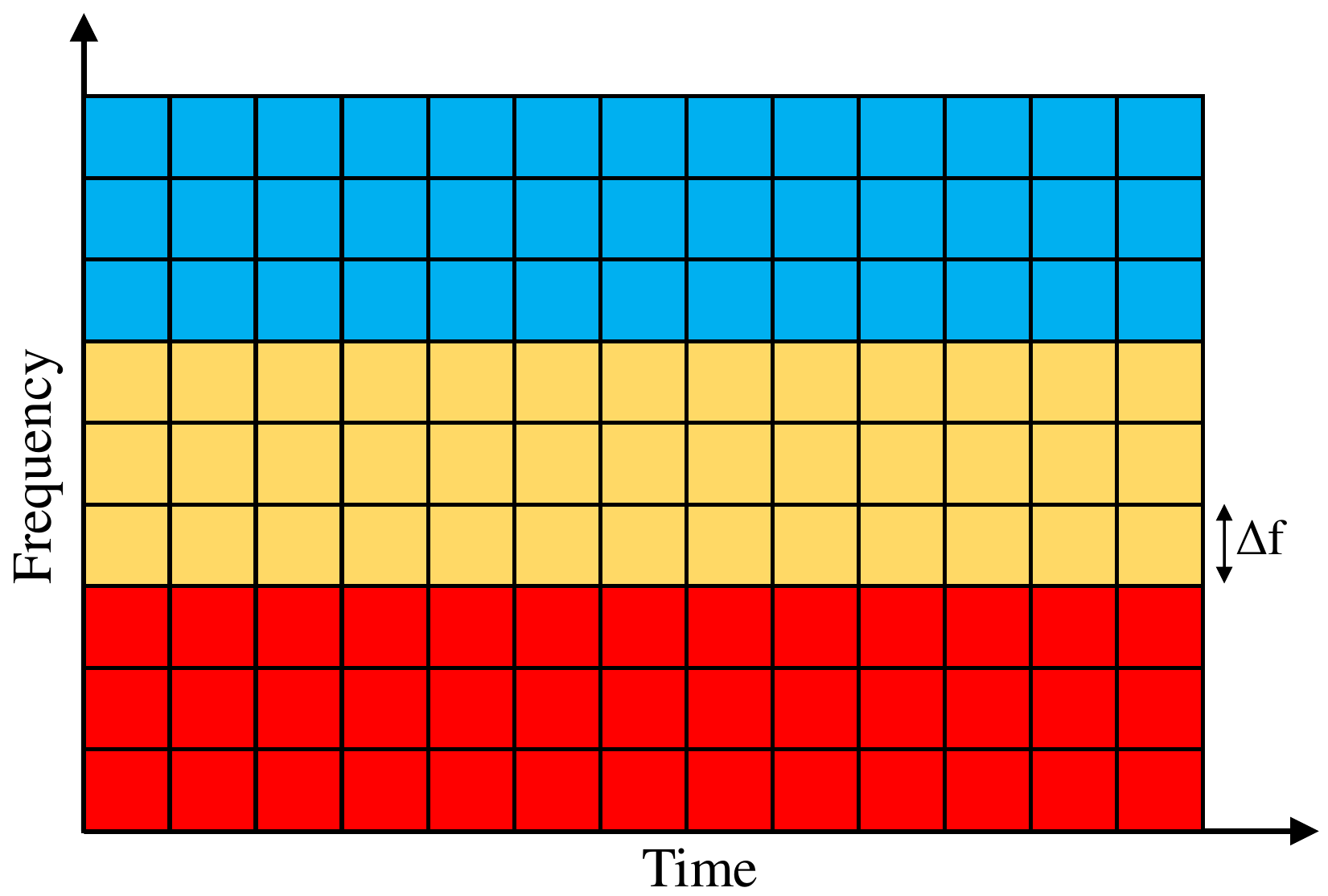} 
  \caption{Block-type multiuser access.}
  \label{sub_block}
\end{figure}

Block-type MU access is conducted by allocating subcarriers to multiple users by groups of subcarrier as depicted in Fig. \ref{sub_block}, e.g. subcarrier 0,1,2 for User 1, subcarrier 3,4,5 for User 2, and subcarrier 6,7,8 for User 3. Inherently, block assignment is convenient for users to manipulate their own transmission in their assigned blocks of subcarriers. However, if channels suffer from notorious selective frequency fading, certain frequency bands may not be able to communicate efficiently. Therefore, this is the motivation for us to exploit alternative MU access scenario.

\subsubsection{Comb-type MU Access}
Subcarriers can also be allocated that each user's subcarriers are distributed uniformly among subcarriers in frequency domain. Hence, in deep fading cases, ineffective frequency bands can distribute among users instead of deteriorating certain user as in block-type MU access. However, the scenario has its drawbacks as well, such as each user demand to manage their information symbols among frequency subcarriers. In this situation, received symbols may suffer from inter symbol interference (ISI) among multiple users.

Therefore, our problem objective is MU-MCDM systems can adapt to choose the parameters, including number of users, number of subcarriers, portion of pilot subcarriers, and MU scenarios in underwater multipath environments. Adaptation can make the system adjust its transmit strategies according to channel's conditions to enhance efficiency in communications. In addition, receiver design comprises packet synchronization, channel estimation, and ML symbol detection procedures.

\begin{figure}
 \centering
 \includegraphics[width=0.8\textwidth]{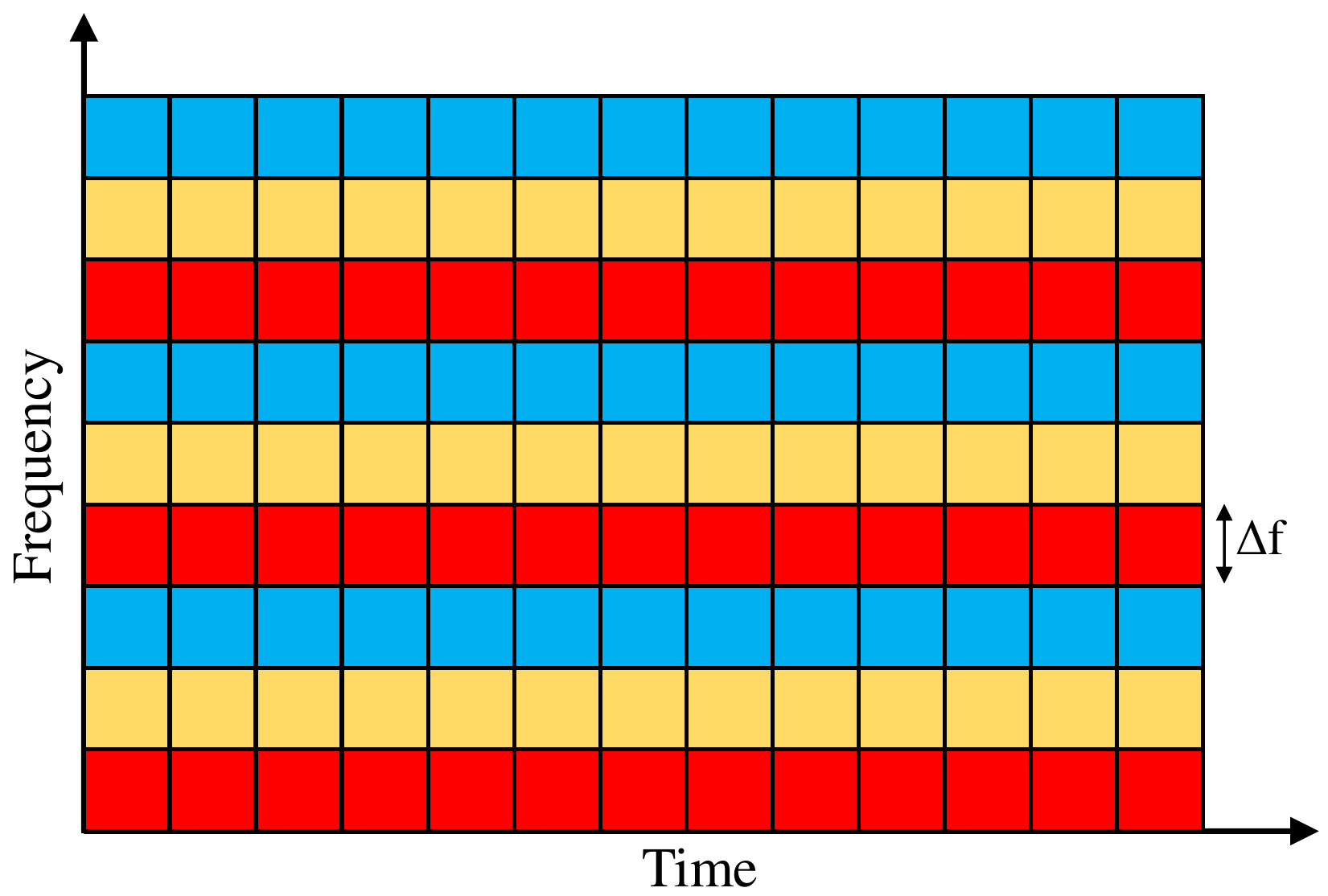} 
  \caption{Comb-type multiuser access.}
  \label{sub_comb}
\end{figure}

\section{Receiver Design}
\label{S4}

\begin{figure}
 \centering
 \includegraphics[width=0.8\textwidth]{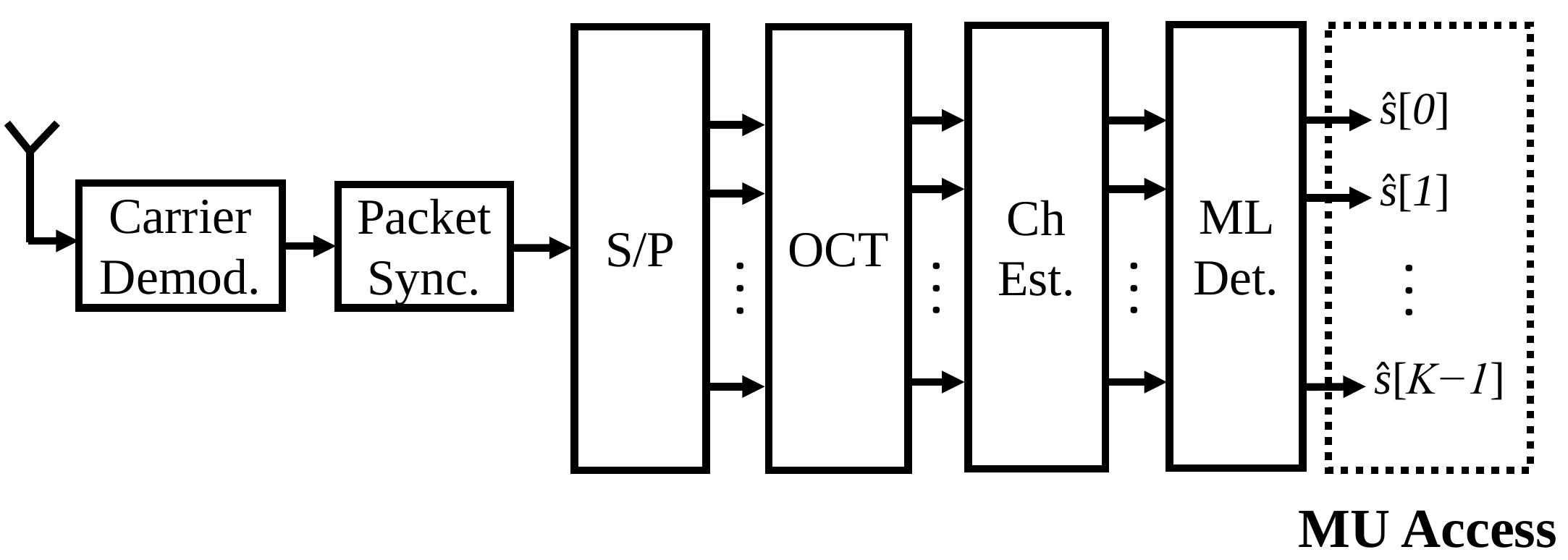} 
  \caption{MU-MCDM receiver system model.}
  \label{Sys_Rx}
\end{figure}

Receiver design consists of packet synchronization, channel estimation, and symbol detection illustrated in Fig. \ref{Sys_Rx}. The signal processing is in baseband after carrier frequency demodulation. OCT is conducted to inverse IOCT procedure in the transmitter side. Then, channel estimation and ML detection is processed in frequency.

\subsection{Packet Synchronization}
First, received signals are conducted by carrier frequency demodulation over a time window. Preambles are utilized for packet synchronization, containing $N_{pn}$ antipodal bits in time duration $T_{pn}$. Considered window should at least cover the received preamble symbols for the synchronization purpose.

To synchronize a received packet with a transmitted packet, correlation of transmitted preambles with received signals is written as
\begin{align}
\label{auto}
R_{pn}(\tau) = \int_0^{T_{pn}} x_{pn}(t)r^*(t+\tau)dt,
\end{align}
where $r(t+\tau)$ is the received signal time shifted by $\tau$, $x_{pn}(t)$ is transmitted preamble signal in time, and $\tau \in [0,~T_w]$.

So, the packet is synchronized at the maximum of square L2-norm of the correlated output described as
\begin{align}
\label{k_syn}
{\hat{t}_r} = arg\ \max_{\tau \in [0,~T_w]} ||R_{pn}(\tau)||^2_2.
\end{align}
Consequently, the received packet is synchronization at the time $\hat{t}_r$, so the beginning of a packet's transmission is located.

\subsection{Channel Estimation}
Then, OCT is conducted for analyzing signals in frequency domain. For the $k$-th subcarrier, OCT is conducted in frequency band $\in [k\Delta f,~k\Delta f+B_c]$. Moreover, MCDM symbols are sampled at the time duration $T_s = \frac{1}{f_s}$. The OCT procedure in the receiver is to inverse the IOCT process in the transmitter. The receiver MCDM vector after carrier frequency demodulation can be expressed as
\begin{align}
\label{y_base}
\mathbf{y} = \mathbf{H} \mathbf{s} + \mathbf{n},
\end{align}
where $\mathbf{y} \in \mathbb{C}^{K \times 1}$ is the received vector, $\mathbf{H}\in \mathbb{C}^{K \times K}$ the channel matrix formed as,
\begin{align}
{\mathbf H} \overset{\Delta}{=} \left[\begin{array}{ccccc}
       h_0    & 0      & \cdots  & 0       & 0\\
       0      & h_1    & \cdots  & 0       & 0\\
       \vdots &\vdots  & \ddots  & \vdots  & \vdots\\
       0      & 0      & \cdots  & h_{K-2} & 0\\
       0      & 0      & \cdots  & 0       & h_{K-1}
     \end{array}\right]
\label{eq:6}
\end{align}
$\mathbf{s} \in \mathbb{C}^{K \times 1}$ the transmitted symbol vector, $\mathbf{n}\in \mathbb{C}^{K \times 1}$ the received noise vector, and $K$ is the overall number of frequency subcarriers. 

Channel model in a diagonal matrix is built on the assumption that bandwidth of a subcarrier lie within channel's coherence bandwidth \cite{coulson01}, whereas a channel's amplitude of a subcarrier can be redeemed as unchanged in a subcarrier's duration. To be consistent to CSI in our system model, the channel coefficient of a subcarrier should be estimated as a complex number $\in \mathbf{C}$. Therefore, the computational overhead of following channel estimation and symbol detection decreases dramatically in MCDM systems.

Pilot symbols are arranged in comb-type subcarrier allocation \cite{shen06}, which pilots are spread evenly among subcarriers and subcarriers' allocation does not change over time. Total number of pilot symbols $K_p$, they are located at $k$ = $0$, $L$, $2L$, ~\ldots~,$(K_p-1)L$, where $L \overset{\Delta}{=} \frac{K}{K_p}$. Therefore, channel states are estimated every $L$ subcarrier for capturing CSI effectively. 

Then, the received pilot vector in baseband can be described as
\begin{align}
\label{y_pilot}
\mathbf{y}_{\it p} = \mathbf{S}_{\it p} \mathbf{h}_{\it p} + \mathbf{n}_{\it p},
\end{align}
where $\mathbf{y}_p \in \mathbb{C}^{K_p \times 1}$ is the received pilot vector, $\mathbf{h}_{\it p} = [h_{p,0},\ h_{p,1},\ \cdots,\ h_{p,K_p-1}]^T \in \mathbb{C}^{K_p \times 1}$ the pilot channel vector, $\mathbf{S}_{\it p}\in \mathbb{C}^{K_p \times K_p}$ the pilot symbols in a diagonal matrix form,
\begin{align}
{\mathbf S_p} \overset{\Delta}{=} \left[\begin{array}{ccccc}
       s_{p,0} &0       &\cdots  &0\\
       0       &s_{p,1}&\cdots   &0\\
       \vdots  &\vdots  &\ddots  &\vdots\\
       0       &0       &\cdots  &s_{p,K_p-1}
     \end{array}\right]
\label{s_p}
\end{align}
and $\mathbf{n}_{\it p} \in \mathbb{C}^{K_p \times 1}$ is the pilot noise vector.

Moreover, channel estimation for pilot symbols can be written as
\begin{align}
\label{pro_che}
\mathbf{\widehat{h}_{\it p}} &= arg\ \min_{\mathbf{h_p} \in \mathbb{C}^{K_p \times 1}} ||\mathbf{y}_{\it p} - \mathbf{S}_{\it p} \mathbf{h}_{\it p}||^2_2.
\end{align}

Pilot matrix $\mathbf{S}_{\it p}$ is known a-priori. So under additive white Gaussian noise (AWGN) condition, ML-optimal channel estimator \cite{Proakis07} can be expressed as
\begin{align}
\label{h_ML}
\mathbf{\widehat{h}_{\it p,ML}} &= (\mathbf{S}_p^H \mathbf{S}_p)^{-1}\mathbf{S}_p^H \mathbf{y}_p.
\end{align} 

ML channel vector for pilot symbols is estimated and we make assumption that a subcarrier's channel is narrowband and channel coefficients vary linearly between neighbored subcarriers. Hence, linear interpolation is conducted for channel coefficients for the non-pilot subcarrier $k$ as 
\begin{align}
\label{h_inter}
\widehat{h}_{ML}[k] = \left(1-\frac{l}{L}\right)\widehat{h}_{p,ML}[m] + \frac{l}{L}\widehat{h}_{p,ML}[m+1],
\end{align}
where $\widehat{h}_{p,ML}[m]$ is the estimated channel coefficient of the $m$-th pilot symbol, $l$ the residue of $k$ divided by $L$, $mL < k < (m+1)L$, and $m = 0,\cdots, K_p-1$. 

Consequently, channel coefficients of all the frequency subcarriers are estimated, so symbol detection can be further processed.

\subsection{ML Symbol Detection}
Our detection objective is to minimize the error between received vector $\mathbf{y}$ and channel-filtered vector $\mathbf{Hs}$ described as
\begin{align}
\label{prob}
\mathbf{\widehat{s}} &= arg\ \min_{\mathbf{s} \in \mathbb{C}^{K \times 1}} ||\mathbf{y} - \mathbf{H} \mathbf{s}||^2_2.
\end{align}

In AWGN, the ML-optimal detection for (\ref{prob}) can be written in a closed-form expression as 
\begin{align}
\label{s_ML}
\mathbf{\widehat{s}_{\it ML}} &= (\mathbf{H}^H \mathbf{H})^{-1} \mathbf{H}^H \mathbf{y},
\end{align} 
where $\mathbf{H}$ is estimated by channel coefficients in (\ref{h_ML}) and (\ref{h_inter}), and then formulated in a diagonal matrix as (\ref{eq:6}). 

As a result, a low complexity receiver design is proposed, including packet synchronization, OCT process, channel estimation, and ML symbol detection. Then, MU-MCDM systems' performance demands to be evaluated.

\section{Discussion}
\label{S5}
Several implementations for our proposed approaches are discussed and analyzed. OCT can enable continuous-time practices, symbol transmission, and channel model in time-domain. On the other hand, DOCT can let us to proceed digital practices and signal processing in discrete manner. More details are described in th following.

\subsection{Relationships of continuous and discrete OCT}
Traditionally, OCT is defined as implementations and applications utilized in continuous-time and system design. Moreover, for transmitter system model or antenna design, they are all time-domain procedures, so we can handle them with OCT and other continuous-time operations. On the other hand, various practical implementations are in need in discrete or digital forms, e.g., digital signal processing, so we develop DOCT for the demand of correspondent digital applications. 

\subsection{Compatibility to OFDM Architectures}
MCDM systems are compatible with prevailing OFDM architectures, so this let our approaches as promising and alternative scenarios for applying in current dominant multicarrier systems. We only need to include the frequency sweeping term, i.e., chirped signals, for each frequency subcarrier as illustrated in Fig.. 

\section{Simulation Studies}
\label{S6}

\begin{figure}
 \centering
 \includegraphics[width=0.8\textwidth]{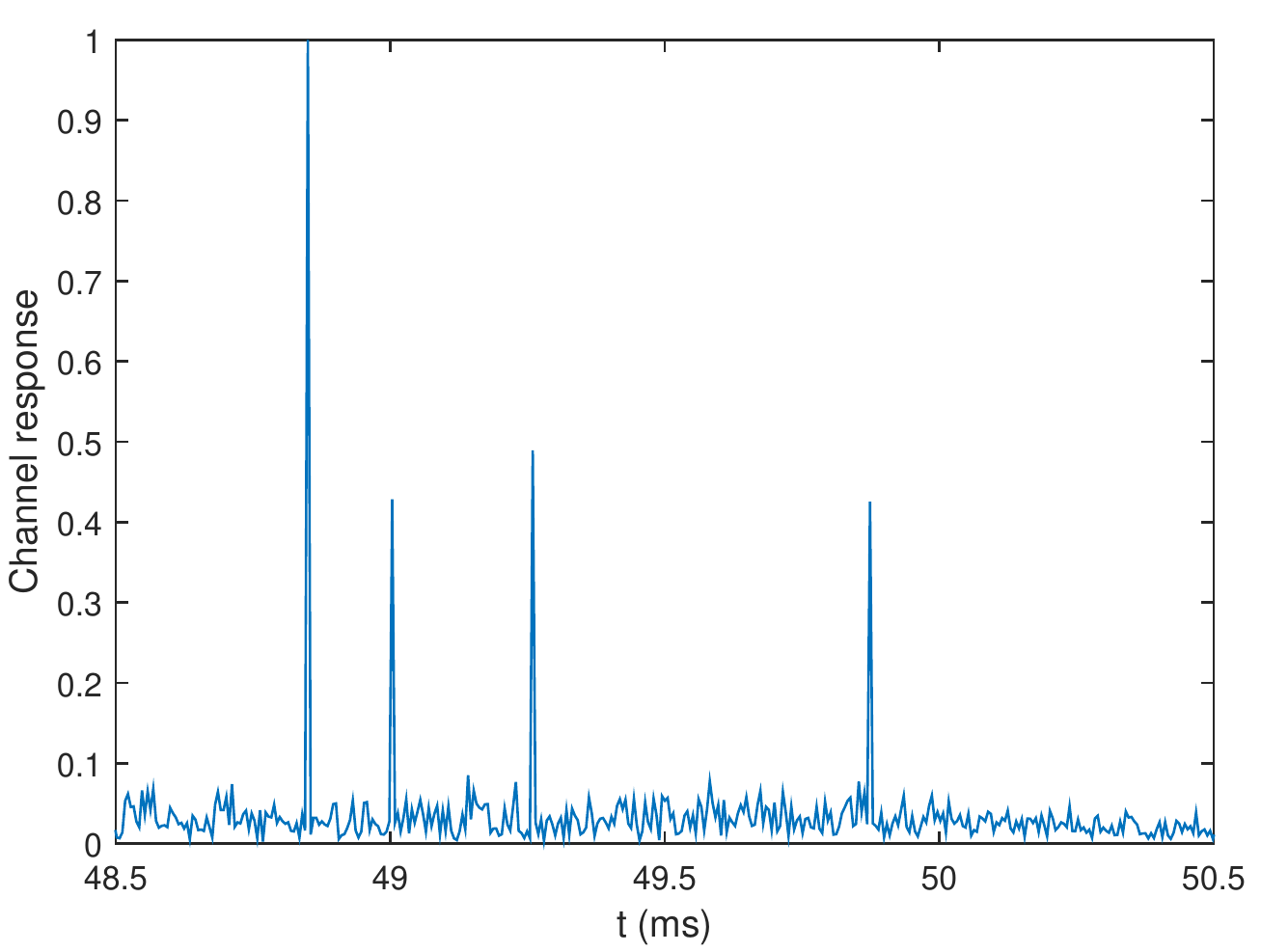} 
  \caption{Normalized channel response in simulations.}
  \label{ch_Sim} 
\end{figure}

In simulations, MCDM systems are simulated in UW-A multipath emulators. Channel's amplitude is modeled as Rayleigh distribution; noise is AWGN, and 4 resolvable paths are shown in Fig. \ref{ch_Sim}. We consider several modulations, including binary phase shift keying (BPSK) and quadrature PSK (QPSK). Moreover, bit mapping for symbols is gray coded for reducing detection errors in adjacent symbols. Total number of subcarriers $K$ can be adaptive to choose from 128, 256, 512, and 1024; portions of pilot subcarriers are available for $\frac{1}{32}$, $\frac{1}{16}$, $\frac{1}{8}$, and $\frac{1}{4}$. Carrier frequency is $f_c$ = 100 kHz. The frequency spacing between subcarriers $\Delta f$ can be 64, 128, 255, 509 Hz. The MCDM symbol period $T$ can be 1.97, 3.93, 7.86, and 15.73 ms, the preamble duration $T_{pn}$ = 1.31 ms, the pause interval between preambles and a MCDM symbol $T_p$ = 1.54 ms, and the guard interval between MCDM symbols $T_g$ = 2.56 ms. Linear up-chirps are applied, chirp rate $\mu$ can be $8.94 \times 10^3$, $1.79 \times 10^4$, $3.58 \times 10^4$, $7.15 \times 10^4$ Hz/s. Available bandwidth for the MCDM system is 65.25 kHz. The simulation results are averaged over 1000 recorded packets.

\begin{figure}
 \centering
 \includegraphics[width=0.8\textwidth]{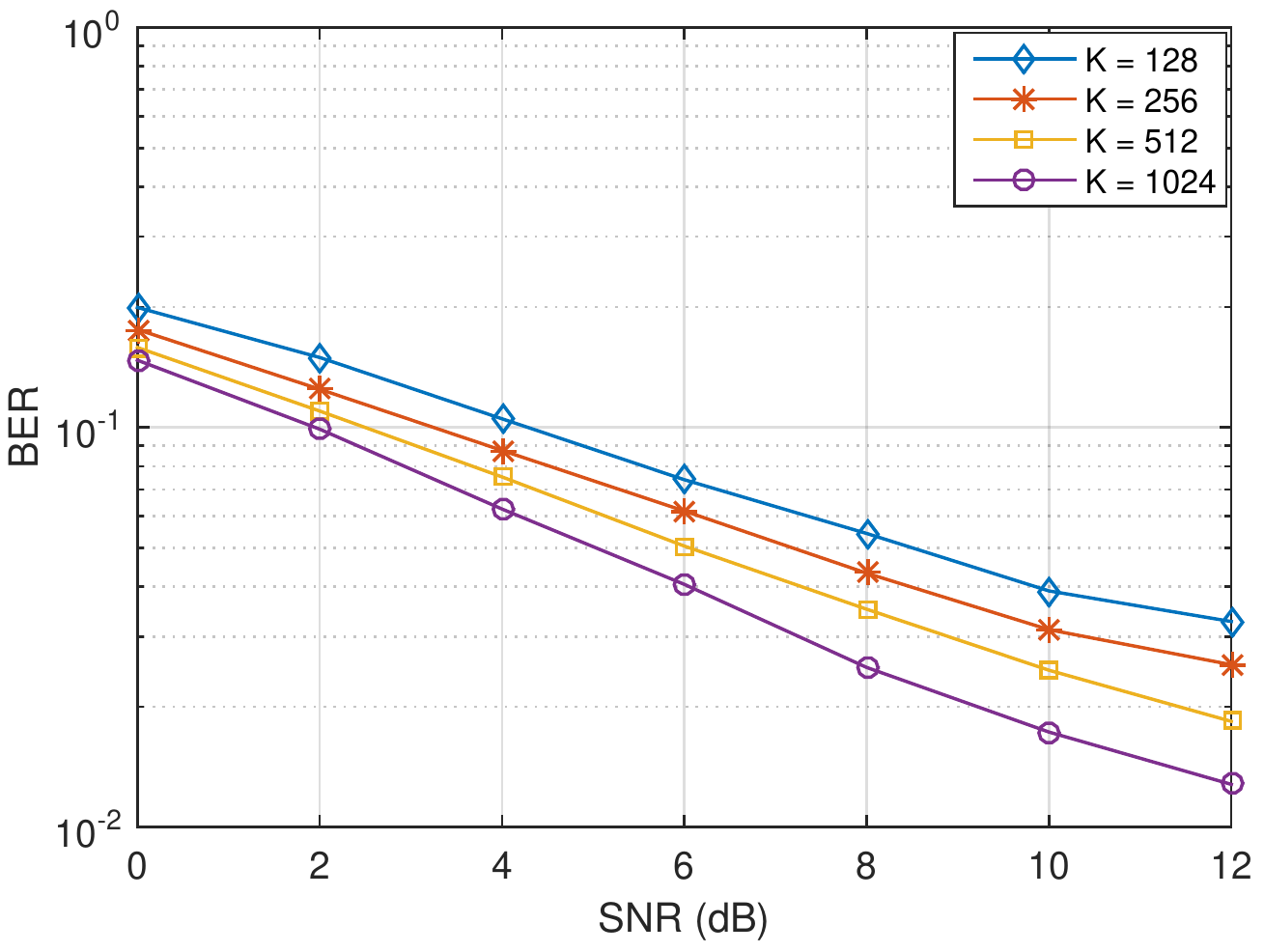} 
  \caption{MCDM for different number of subcarriers.}
  \label{Sim_K}
\end{figure}


Fig. \ref{Sim_K} shows BER vs. SNRs for different number of subcarriers $K$. The portion of subcarriers is fixed at $\frac{1}{4}$ for different $K$. BER indicated here is the average BER for all users (URs). If we compare $K$ = 1024 with $K$ = 512, BER is decreased for 1.58 dB at SNR = 12 dB. The more the number of subcarrier, the lower the BER in MCDM systems. Among the available number of subcarriers, BER can be reach to 1.28 $\times 10^{-2}$ at $K$ = 1024 at SNR = 12 dB.



\begin{figure}
 \centering
  \includegraphics[width=0.8\textwidth]{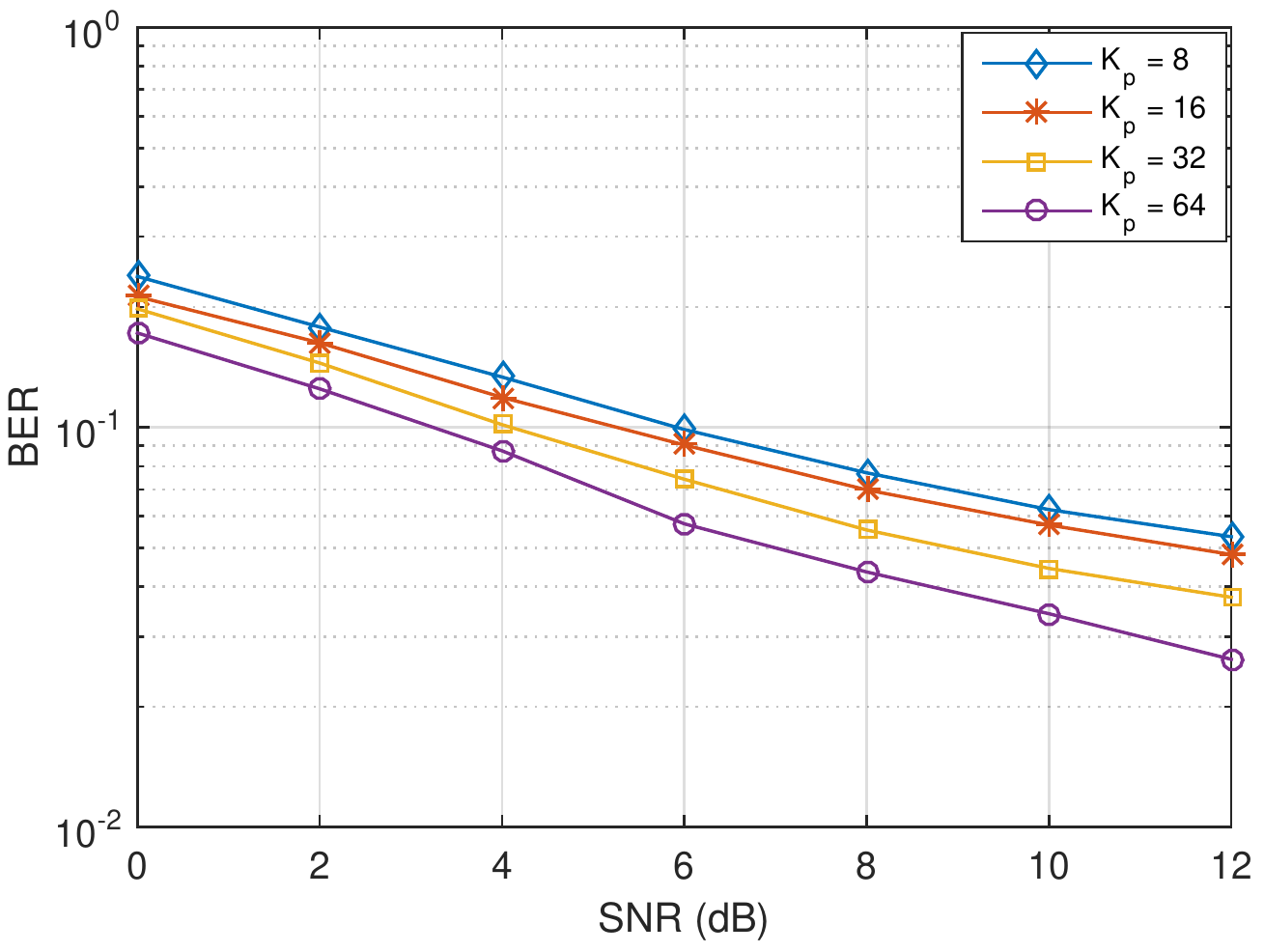} 
  \caption{K = 256 for different number of pilots.}
  \label{Sim_256_Kp}
\end{figure}


In Fig. \ref{Sim_256_Kp} illustrates BER to SNR values of different pilots' portion for $K$ = 256. BER improvement for $K_p$ = 16 and $K_p$ = 8 is 0.45 dB, while that for $K_p$ = 64 and $K_p$ = 32 is 1.55 dB. BER decrease between adjacent curves is larger for higher $K_p$. For fixed $K$, as the number of pilots increases, respective BER is reduced significantly since CSI can estimated more accurately by increasing number of pilot symbols.

\begin{figure}
 \centering
  \includegraphics[width=0.8\textwidth]{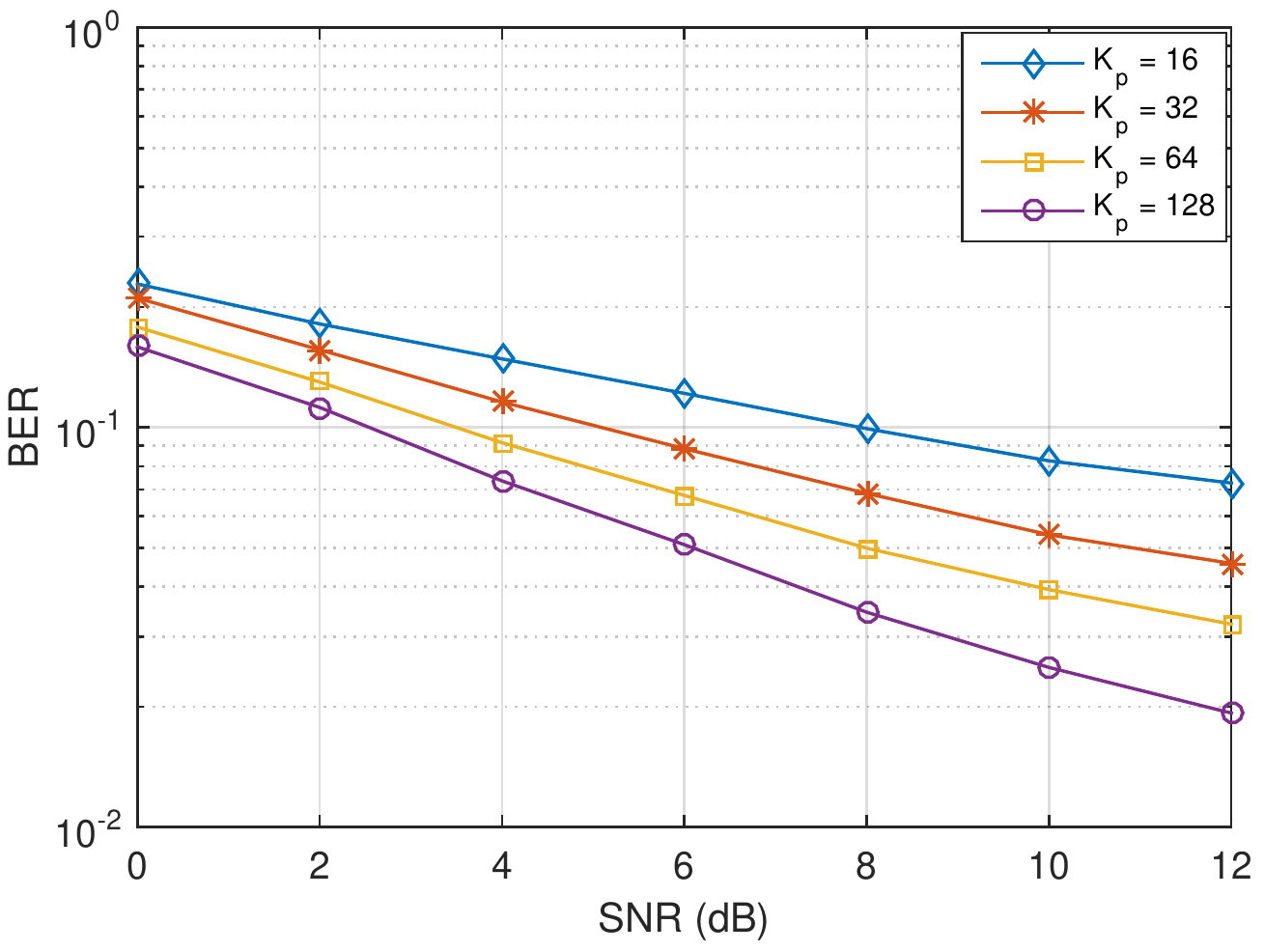} 
  \caption{K = 512 for different number of pilots.}
  \label{Sim_512_Kp}
\end{figure}

\begin{figure}
 \centering
 \includegraphics[width=0.8\textwidth]{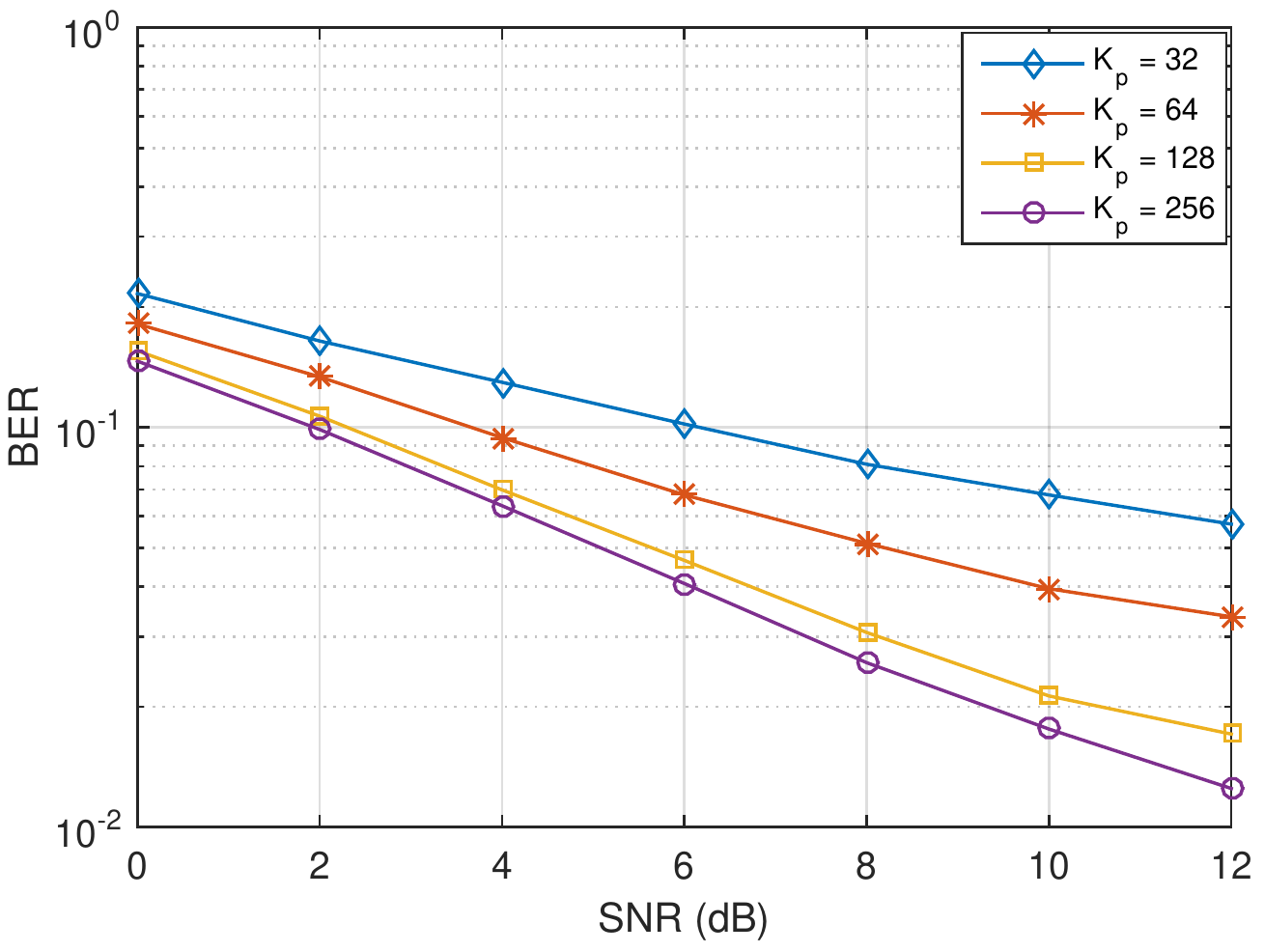} 
  \caption{K = 1024 for different number of pilots.}
  \label{Sim_1024_Kp}
\end{figure}

\begin{figure}
 \centering
 \includegraphics[width=0.8\textwidth]{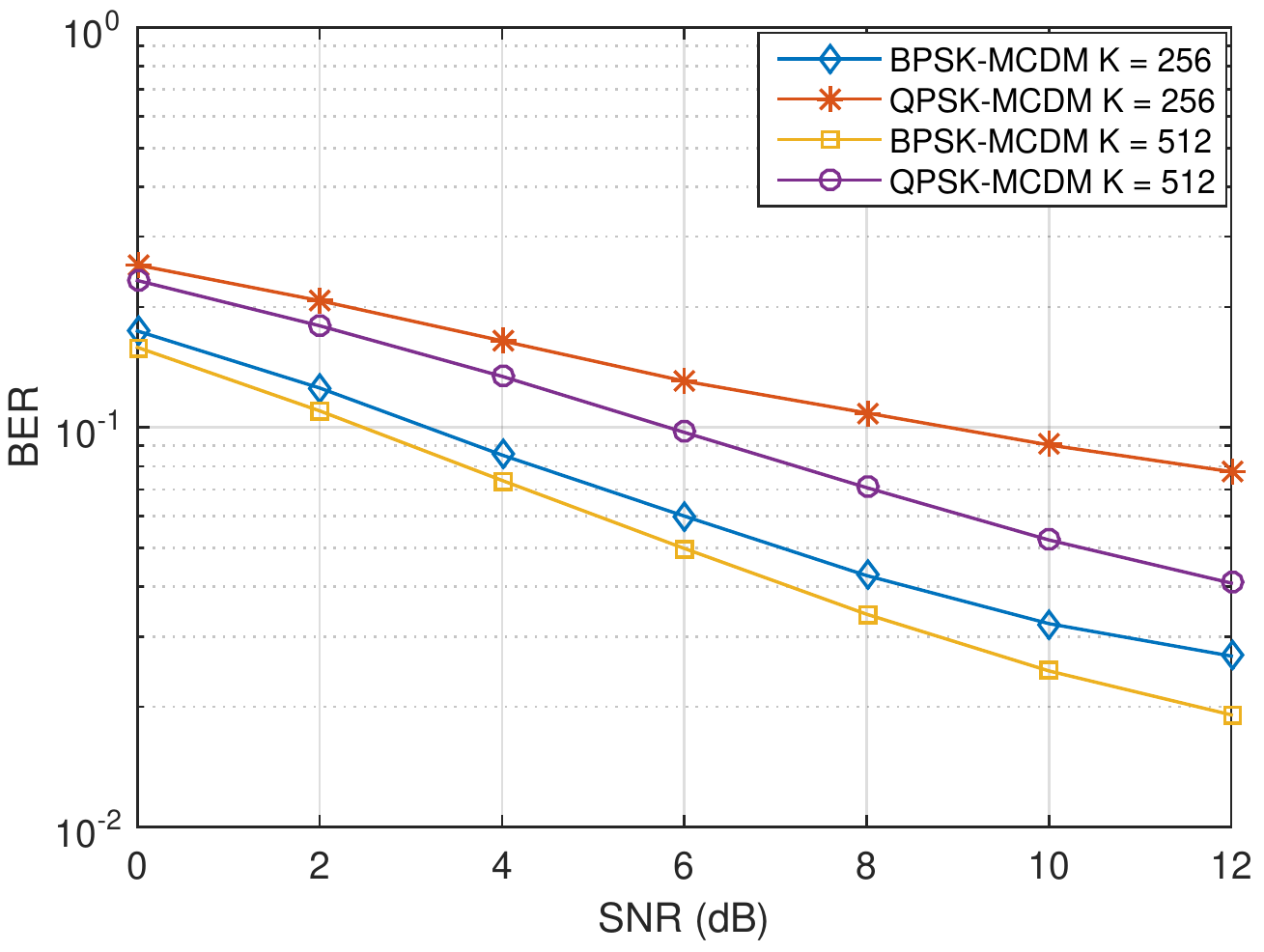} 
  \caption{PSK-MCDM for different number of subcarriers.}
  \label{Sim_PSK_K}
\end{figure}

Fig. \ref{Sim_512_Kp} shows BER of different pilots' portion for $K$ = 512. BER of $K$ = 128 is improved for 1.95 dB to that of $K$ = 64 at SNR = 10 dB, whereas BER improvement for $K$ = 64 and $K$ = 128 is 1.52 dB at SNR = 12 dB. In the $K$ = 512 case, BER follows the trend as well that BER is reduced for larger number of pilots.

Fig. \ref{Sim_1024_Kp} demonstrates BER of different pilots' portion for $K$ = 1024. As the same as previous simulation results, higher number of pilot symbols can benefit MCDM systems in BER performance. Moreover, at SNR = 10 dB, BER reduction of $K_p$ = 128 to that of $K_p$ = 256 is 0.83 dB, which is smaller than that of $K_p$ = 64 to BER of $K_p$ = 128 2.68 dB. I suppose the rationale is that average CSI variants among subcarriers is around 4 to 8 subcarrier interval. Hence, $K_p$ = 256 can update channel information every 4 subcarriers, which some updates of channel coefficients cannot benefit MCDM systems as expected in detections.


\begin{figure}
 \centering
 \includegraphics[width=0.8\textwidth]{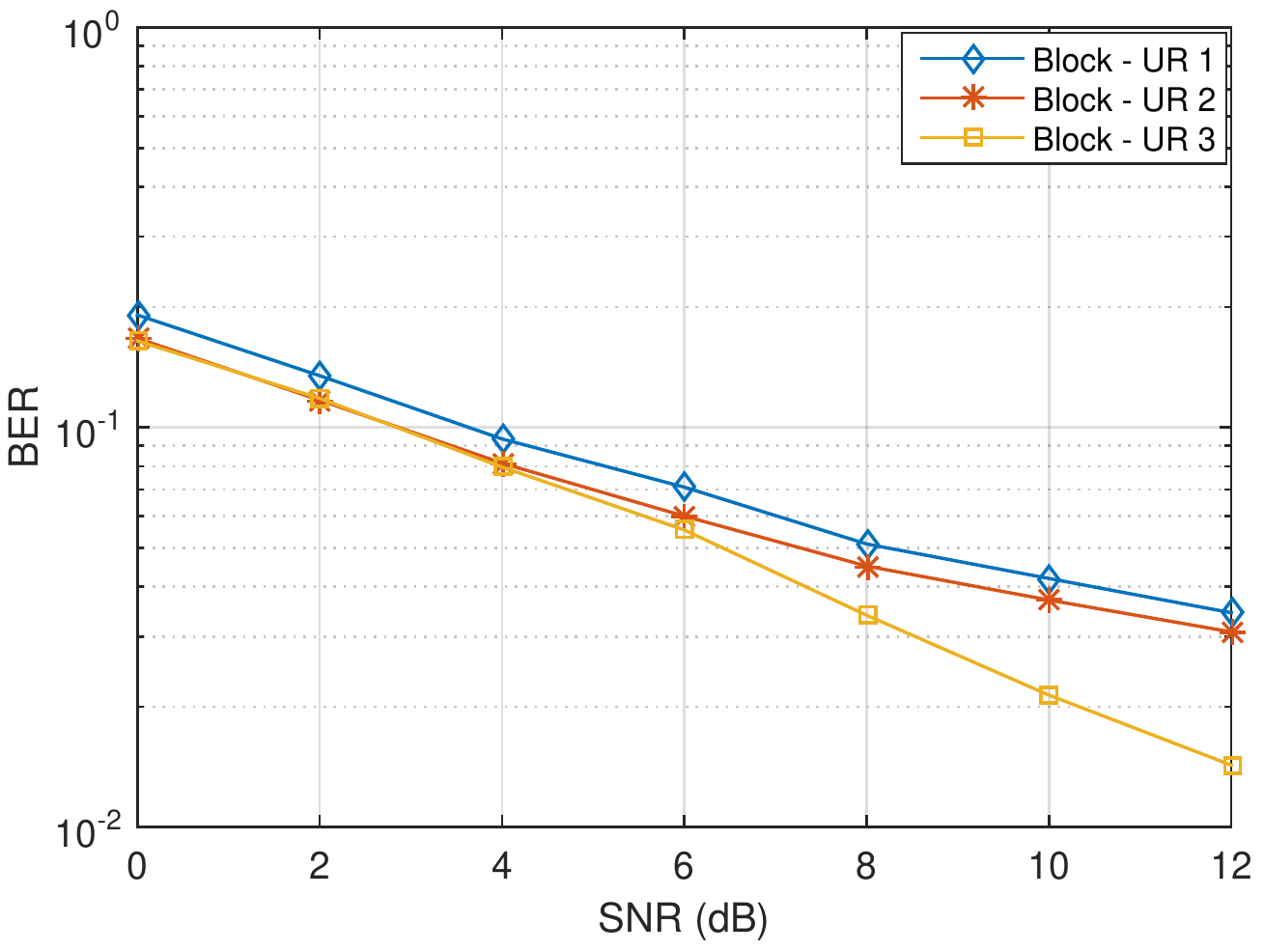} 
  \caption{Block-type access with $N_{UR}$ = 3.}
  \label{Sim_block_3}
\end{figure}

\begin{figure}
 \centering
 \includegraphics[width=0.8\textwidth]{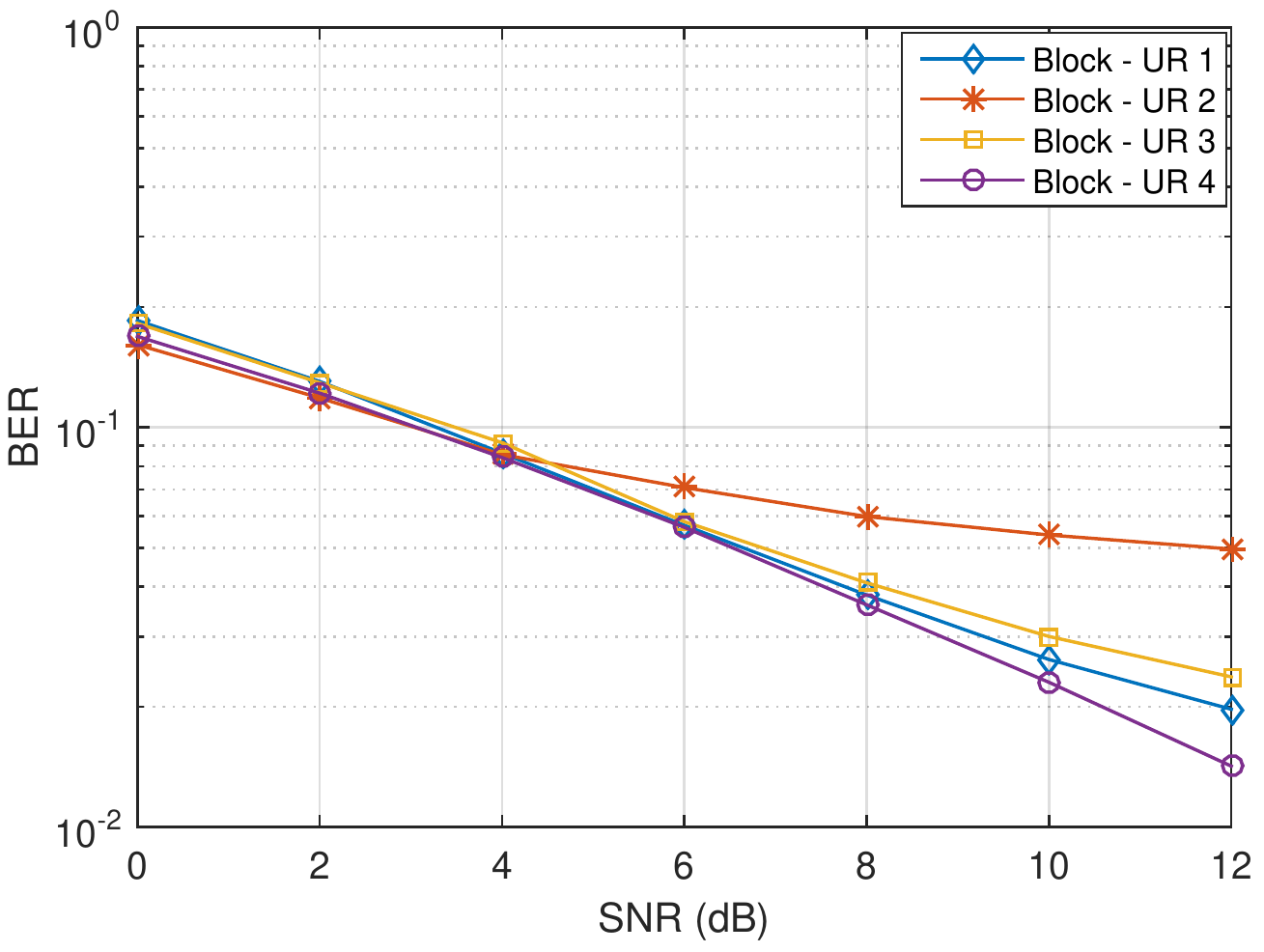} 
  \caption{Block-type access with $N_{UR}$ = 4.}
  \label{Sim_block_4}
\end{figure}

\begin{figure}
 \centering
 \includegraphics[width=0.8\textwidth]{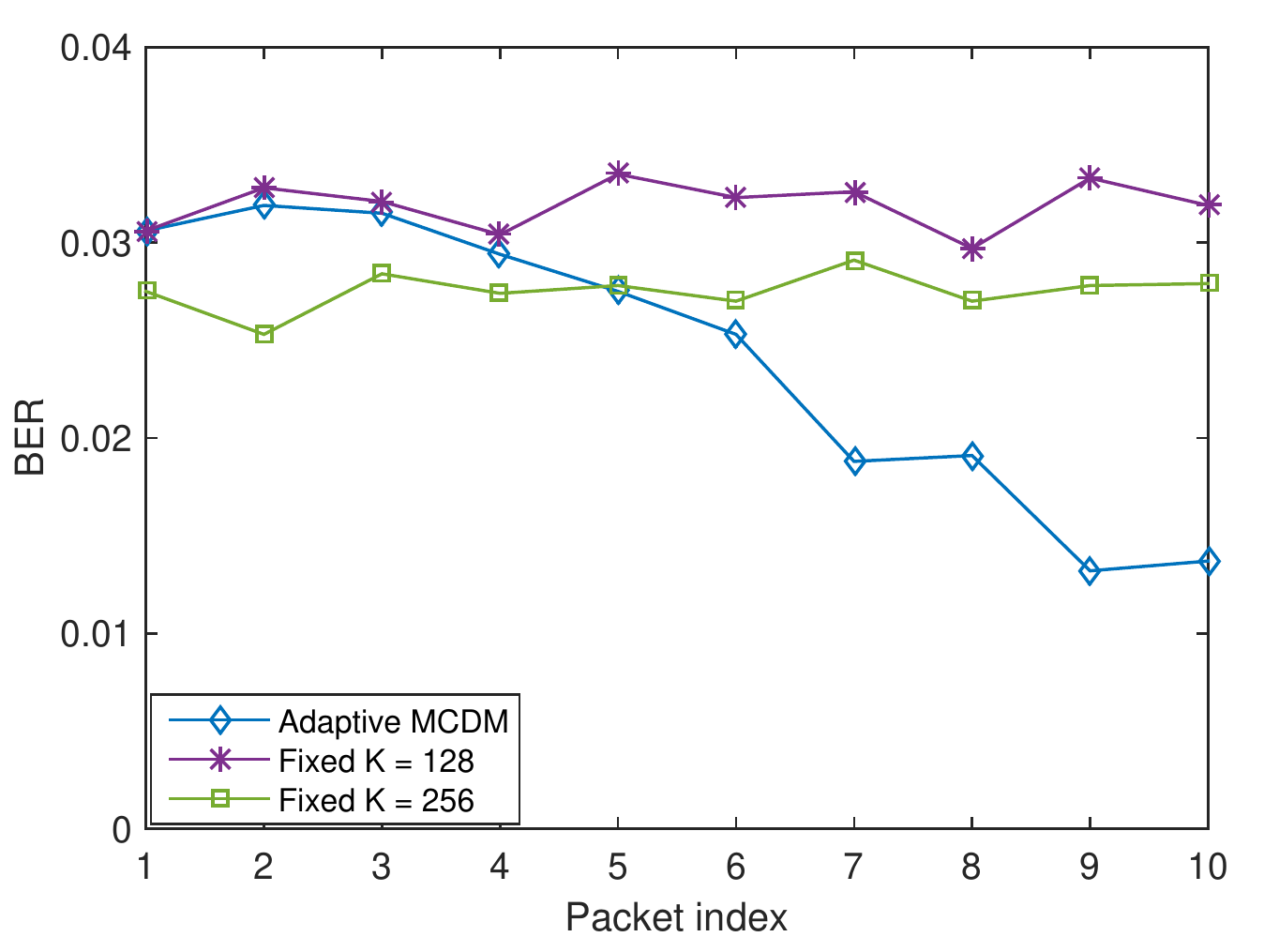} 
  \caption{Adaptive MCDM in block-type access.}
  \label{Sim_block_adap}
\end{figure}

\begin{figure}
 \centering
 \includegraphics[width=0.8\textwidth]{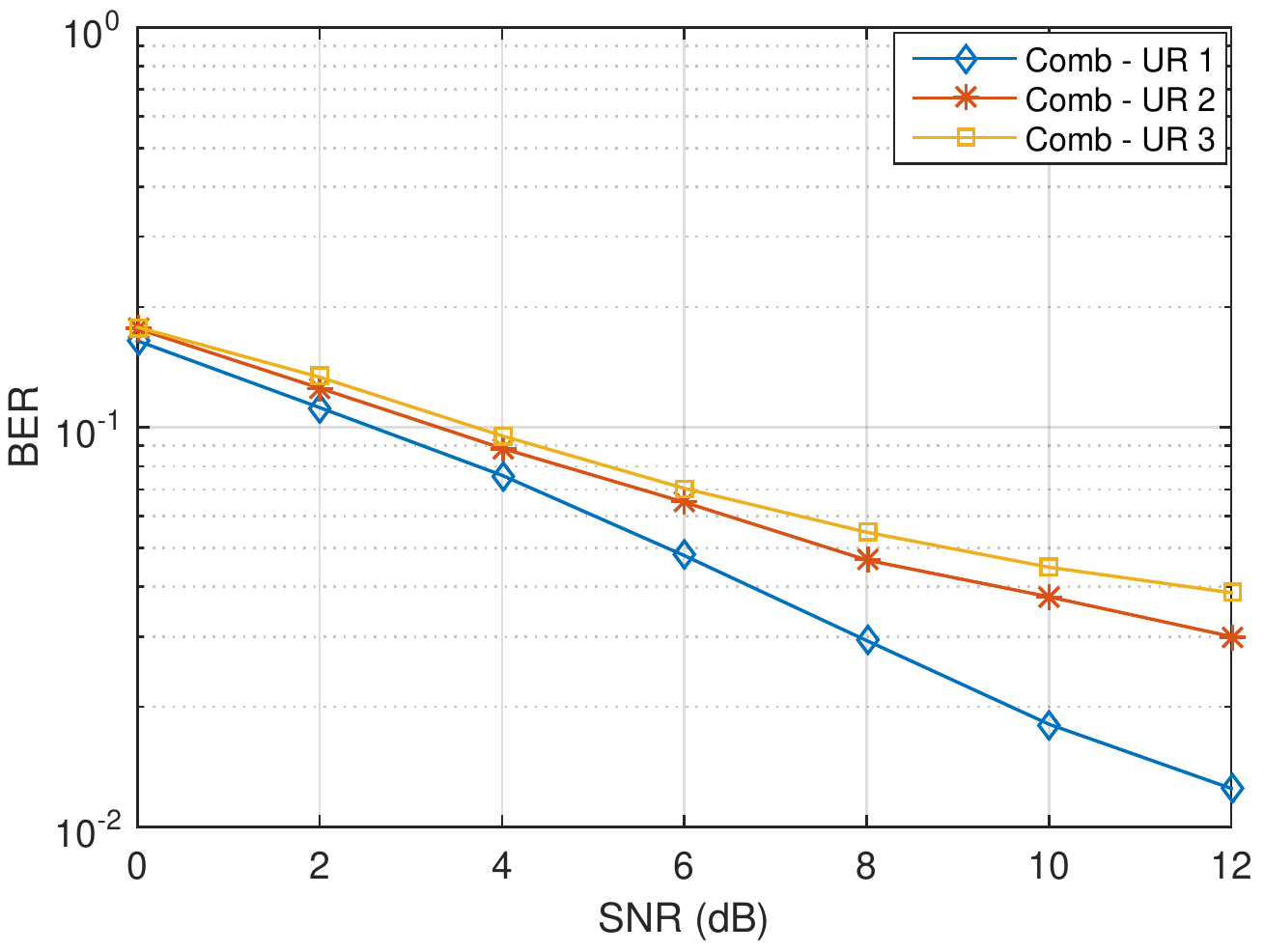} 
  \caption{Comb-type access with $N_{UR}$ = 3.}
  \label{Sim_comb_3}
\end{figure}

\begin{figure}
 \centering
 \includegraphics[width=0.8\textwidth]{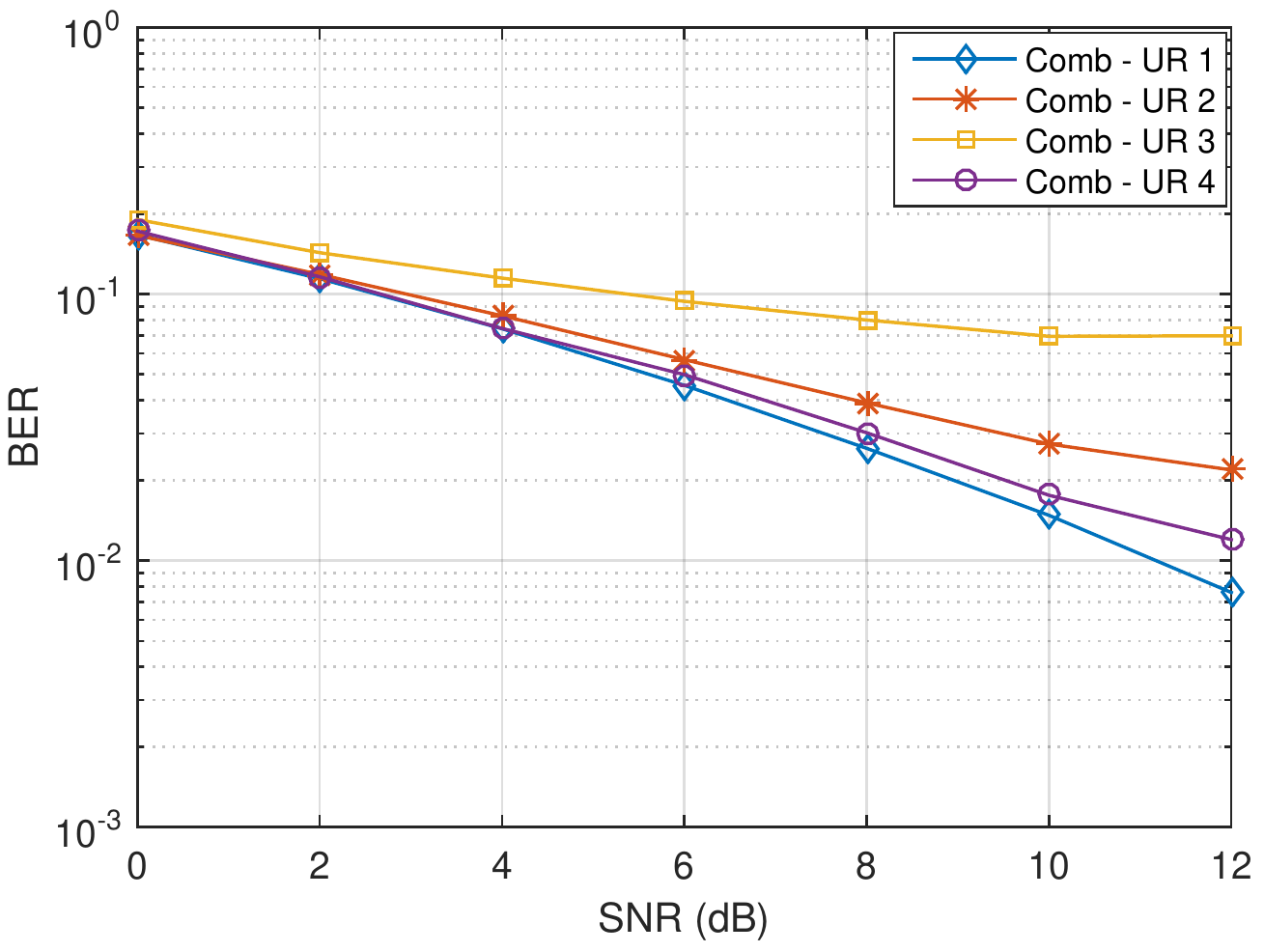} 
  \caption{Comb-type access with $N_{UR}$ = 4.}
  \label{Sim_comb_4}
\end{figure}

\begin{figure}
 \centering
 \includegraphics[width=0.8\textwidth]{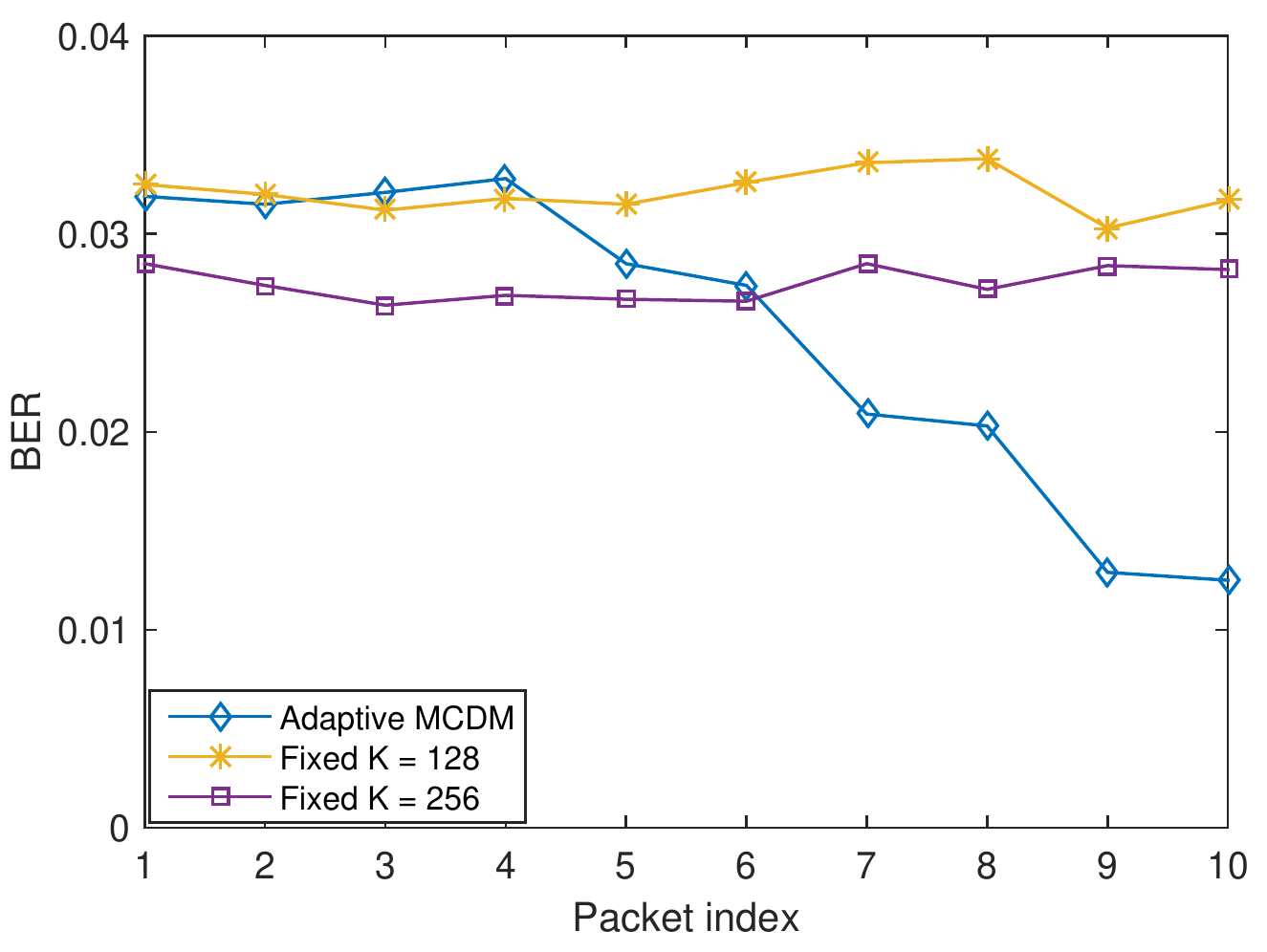} 
  \caption{Adaptive MCDM in comb-type access.}
  \label{Sim_comb_adap}
\end{figure}

Fig. \ref{Sim_PSK_K} demonstrates BER of PSK-MCDM for $K$ = 256 and 512. If we compare BPSK, $K$ = 512 can decrease BER for 1.18 dB to that of $K$ = 256 at SNR = 10 dB. Also, for $K$ = 512, QPSK can increase BER for 3.30 dB than BPSK at SNR = 12 dB, while bit rate of QPSK can be two times to that of BPSK. Hence, it is a trade-off for BER and transmission rate in designing MCDM systems. All in all, higher number of subcarrriers can enhance BER performance in MCDM systems.

Fig. \ref{Sim_block_3} illustrates BER vs. SNR for $N_{UR}$ = 3 for $K$ = 256 in block-type MU access. Subcarriers are allocated as first one-third symbol carriers to UR 1, second one third to UR 2, and third one third to UR 3. The subcarrier allocation is fixed for 1000 packets' transmission. As we can observe from the figure, UR 1 and UR 2 perform worse in BER, while UR 3 can achieve lower BER values. This can be explained as in the multipath Rayleigh fading channel's realizations, UR 3's block of subcarriers suffer less from frequency fading, so it can reach lower BER values in detections. 

Fig. \ref{Sim_block_4} shows BER vs. SNR values for $N_{UR}$ = 4 for $K$ = 256 in block-type MU access. Here, subcarriers are divided into four blocks and assigned to UR 1, 2, 3, and 4, respectively. As we can observe from the figure, UR 1, 3, and 4 can achieve lower BER, while UR 2 obtain worse BER. BER of UR 3 is improved for 3.19 dB to that of UR 2 at SNR = 12 dB. 

Fig. \ref{Sim_block_adap} demonstrates average BER for adaptive MCDM for available $K$ = 128, 256, 512, and 1024 in block-type MU access, SNR fixed at 12 dB. In the practice, there are 10 packets transmitted in multipath fading channels. Packet index 1 to 4, the MCDM system utilize $K$ = 128 for transmission; packet index 5 and 6 for $K$ = 256; index 7 and 8 for $K$ = 512; index 9 and 10 for $K$ = 1024. Based on the simulation results, the MCDM system decides to adopt $K$ = 1024 for future transmission. The feedback information is transmitted to the transmitter through a low-rate and robust chirp signals. Hence, the transmitter can adapt to the scenario enhancing BER performance most.


Fig. \ref{Sim_comb_3} shows BER for $N_{UR}$ = 3 for $K$ = 256 in comb-type MU access. Users' subcarriers are assigned in interleaving manner, e.g., if we have 6 available subcarriers with indexes 0 - 5, subcarrier 0 and 3 for UR 1, subcarrier 1 and 4 for UR 2, subcarrier 2 and 5 for UR 3. Then, the subcarrier allocation is fixed during all the transmission. UR 2 and UR 3 have higher BER values, whereas UR 1's BER is the lowest among the users. During the transmission, UR 1's subcarriers suffer less from channel fading, so its BER values are the best in the simulations.

Fig. \ref{Sim_comb_4} illustrates BER for $N_{UR}$ = 4 for $K$ = 256 in comb-type MU access. Subcarriers are assigned to multiple users by the interleaving procedure. In the practice, UR 2 can decrease BER by 3.11 dB to that of UR 3 at SNR = 8 dB, while BER improvement for UR 1 and UR 4 is merely 0.58 dB. The simulation results are influenced by fading channel characteristics on frequency subcarriers in frequency. If we compare the lowest BER among users in comb-type with block-type MU access, comb-type can improved for 2.71 dB. This can be explained that in flat-fading channels, since comb-type allocation distributes the subcarriers uniformly among users, all the users can be beneficial to their performance.

Fig. \ref{Sim_comb_adap} demonstrates average BER for adaptive MCDM for available $K$ = 128, 256, 512, and 1024 in comb-type MU access, SNR fixed at 12 dB. There are 10 packets transmitted in the simulation. Packet index 1 to 4, the MCDM system utilize $K$ = 128 for transmission; packet index 5 and 6 for $K$ = 256; index 7 and 8 for $K$ = 512; ; index 9 and 10 for $K$ = 1024. Based on the simulation results, the MCDM system decides to adopt $K$ = 1024 for future transmission. Comb-type can decrease BER by 0.24 dB than that of block-type. 


\section{Conclusion}
\label{S7}
In this work, we propose to apply orthogonal chirp waveforms as subcarriers in MCDM systems. Additionally, the proposed communication systems are able to support multiple users' transmission simultaneously. Adaptive transmission strategies are as well considered during the data transmission for optimizing BER performance, including number of subcarriers, number of pilot subcarriers, digital modulations, number of users, approaches to deploy pilot subcarriers. Simulation studies demonstrate that adaptive transmission scenarios outperform other scenarios. The BER performance of the adaptive MU-MCDM is the best among various scenarios. In conclusion, adaptive transmission strategies optimize the performance for MU-MCDM systems.

\bibliographystyle{IEEEtran}
\bibliography{ref}

\end{document}